\documentclass{osa-article}

\usepackage{float}
\usepackage{graphicx,amsmath,ulem}
\usepackage[section]{placeins}
\usepackage{ulem}
\usepackage[export]{adjustbox}[2011/08/13]
\usepackage{verbatim}
\usepackage{ragged2e}
\usepackage{color}
\usepackage{color, colortbl}
\usepackage{physics}

\journal{osac}

\articletype{Research Article}

\begin{document}

\title{Bright correlated twin-beam generation and radiation shaping in high-gain parametric down-conversion with anisotropy}

\author{M. Riabinin,\authormark{1} P. R. Sharapova,\authormark{1} and T. Meier\authormark{1}}

\address{\authormark{1} Department of Physics, University of Paderborn,
Warburger Straße 100, D-33098 Paderborn, Germany}




\begin{abstract}
Uniaxial anisotropy in nonlinear birefringent crystals limits the efficiency of nonlinear optical interactions and breaks the spatial symmetry of light generated in the parametric down-conversion (PDC) process. Therefore, this effect is usually undesirable and must be compensated for. However, high gain may be used to overcome the destructive role of anisotropy and instead to use it for the generation of bright two-mode correlated twin-beams. In this work, we provide a rigorous theoretical description of the spatial properties of bright squeezed light in the presence of strong anisotropy. We investigate a single-crystal and a two-crystal configuration and demonstrate the generation of bright correlated twin-beams in such systems at high gain due to anisotropy. We explore the mode structure of the generated light and show how anisotropy, together with crystal spacing, can be used for radiation shaping.
\end{abstract}

\section{Introduction}

Parametric down-conversion (PDC) is a nonlinear optical process in which the pump field generates pairs of entangled photons inside a nonlinear medium~\cite{PhysRevLett.25.84, PhysRev.124.1646, Karan_2020, WALBORN201087}. The PDC process is an important source of squeezed and correlated light~\cite{PhysRevA.91.043816, PhysRevA.97.053827, PhysRevA.69.023802, PhysRevLett.93.243601}. In birefringent crystals, to realize phase matching of the PDC process, a particular angle between the extraordinary pump wavevector and the crystal optical axis must be set, which results in spatial walk-off of the pump beam inside the crystal. Spatial walk-off is essentially caused by a misalignment between the pump's Poynting vector and wavevector. The angle between the direction of the extraordinary beam outside and inside the crystal defines the walk-off angle~\cite{PhysRevLett.99.063901, Huo_and_Zhang, PhysRevA.77.015805}. As a result, the spatial intensity distribution leans towards the walk-off direction and becomes asymmetric.

Generally, anisotropy is undesirable because, due to the pump's shift, the generated PDC photons quickly leave the interaction region, where they overlap with the pump. This limits the efficiency of nonlinear optical interactions and the use of long crystals, where such effects are more prominent. To eliminate walk-off,
a well-known experimental technique is used, which includes pairs of thin crystals with an oppositely directed optical axis, which allows one to compensate for the spatial asymmetry gained in the first crystal and to increase the interaction efficiency~\cite{P_rez_2013, Cavanna:14}.

However, one can overcome the destructive role of the walk-off effect and even turn it into a useful tool for generating bright narrow twin-beams. For example, in Ref.~\cite{Perez2015} it was shown how the tilt of a crystal (a change in the orientation of the optical axis) leads to the generation of bright separated twin-beams at correlated frequencies, and in Ref.~\cite{P_rez_2013} twin-beams were experimentally generated in a two-crystal configuration with an air gap. Correlated twin-beams~\cite{Hochrainer1508, Rarity1992} are a fundamental resource for quantum metrology~\cite{Pradyumna2020, Clark_2021, Jason_D_Muellera}, quantum imaging~\cite{Brida2010, Celebrano2011}, and spectroscopy~\cite{Whittaker_2017, Losero2018}, as they overcome the shot-noise limit. Recently, more attention has been attracted by bright correlated twin-beams~\cite{Agliati_2005, Li_20, Chekhova_18, Wu_19}, since they include a huge number of photons and are characterized by strong macro correlations.

A theoretical description of the anisotropy effect and its compensation at low gain had been presented in Refs.~\cite{P_rez_2013, Cavanna:14}, where the biphoton amplitude approximation was applied to describe the PDC process. However, this approximation is incomplete for high parametric gains. In Ref.~\cite{PhysRevA.91.043816} the Schmidt-mode approach was used to describe the compensated and non-compensated configurations in the high-gain regime. However, this approach does not provide entirely correct predictions with respect to the mode shapes at high gain~\cite{PhysRevResearch.2.013371}.
 
In this work, we present a detailed theoretical description of bright squeezed light generated in the high-gain PDC process in the presence of strong anisotropy. Our description is based on systems of integro-differential equations for the plane-wave operators and is evaluated for configurations with compensated and non-compensated anisotropy. We demonstrate that at high gain, the anisotropy effect can be utilized as a useful tool for generating two-mode bright spatially correlated twin-beams, even for the collinear phase matching synchronism. Such twin-beams require high gain and cannot be observed in the low intensity regime. However, considering two crystals with an air gap allows such highly correlated beams to be generated at lower pump intensities, while simultaneously the number of generated photons is increased by several orders of magnitude. Finally, we show how anisotropy, together with crystal spacing can be used for radiation shaping and modifying the mode structure at high gain.

\section{A single crystal with anisotropy at high gain}

Here, we investigate spatial properties of light generated in the type-I PDC process in the presence of strong anisotropy and neglect by the frequency distribution of photons by assuming ideal frequency matching  $\omega_{s} \approx 
\omega_{i} \approx \omega_{p}/2$, where the indices 's', 'i', 'p' denote the signal, the idler, and the pump photons, respectively.  For simplicity, we consider a one-dimensional (1D) distribution of the pump field $E_p^{(+)} (\mathbf{r},t)= E_0 e^{-\frac{x^2}{2\sigma_{p}^2}}e^{i( k_p z-\omega_p t)}$ with a Gaussian envelope along the '$x$' axis of full width at half maximum (FWHM) of the intensity distribution being $2\sqrt{\ln 2}\sigma_{p}$ and an amplitude of $E_0$. Such approximation neglects by effects that can be acquired by using arbitrary pump fields~\cite{PhysRevA.81.053805}. As it was shown previously, high-gain properties of generated multimode PDC light can be very well described by a system of integro-differential equations for the plane-wave operators~\cite{PhysRevResearch.2.013371}. In the following, we extend this approach by accounting for the pump walk-off effect inside the crystal due to anisotropy. 

\begin{figure}[H]
  \includegraphics[width=0.5\textwidth, center]{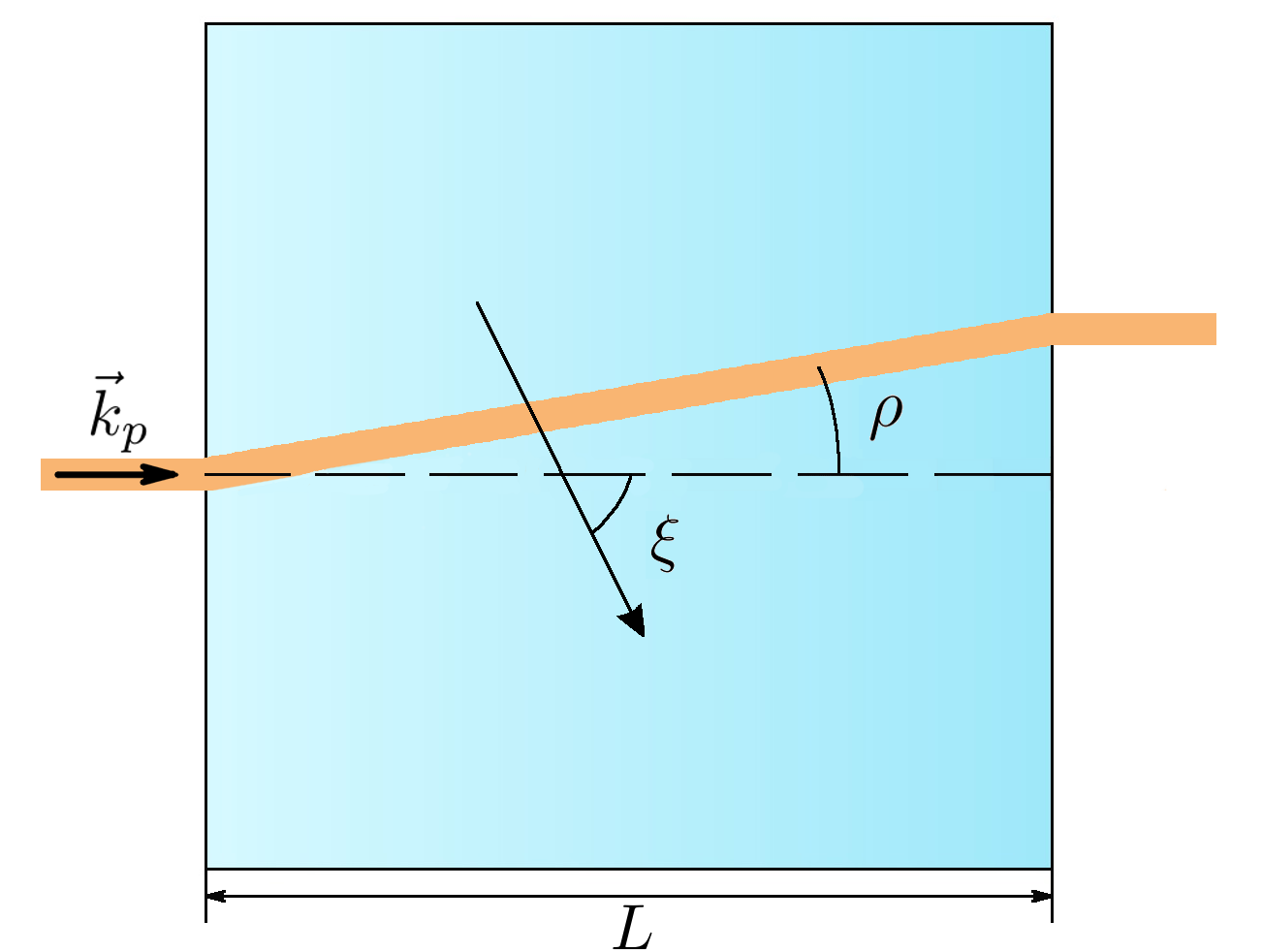}
  \caption{An illustration of the walk-off effect. Inside the birefringent crystal, the extraordinary polarized pump beam deviates from the initial direction. The walk-off angle  $\rho$ is the angle between the initial and the deviated pump directions. The angle $\xi$ is the angle between the crystal's optical axis and the direction of the pump wave vector $\vec{k}_{p}$. 
  }
  \label{fig:one_crystal}
\end{figure}

For a single crystal with anisotropy, the system of integro-differential equations for the plane-wave operators $a_s (q_s, z,\omega_s)$, $a^{\dagger}_i(q_i, z,\omega_i)$ has the form:

\begin{eqnarray}
\frac{\mathrm{d} a_s (q_s, z,\omega_s)}{\mathrm{d}z}=
\Gamma \int \mathrm{d}q_i \exp{-\frac{\sigma_{p}^2}{2}(\Delta k_{\parallel}\sin{\rho} + \Delta k_{\perp}\cos{\rho})^2 } \times \label{eqn:eq1_anis_1cr} \\ \exp{i z (\Delta k_{\parallel} - \Delta k_{\perp}\tan{\rho})} a^{\dagger}_i(q_i, z,\omega_i), \nonumber \ \ \ \ \
\end{eqnarray}
\begin{eqnarray}
\frac{\mathrm{d} a^{\dagger}_i(q_i, z,\omega_i)}{\mathrm{d}z}=
\Gamma \int \mathrm{d}q_s \exp{-\frac{\sigma_{p}^2}{2}(\Delta k_{\parallel}\sin{\rho} + \Delta k_{\perp}\cos{\rho})^2 } \times \label{eqn:eq2_anis_1cr} \\ \exp{-i z (\Delta k_{\parallel} - \Delta k_{\perp}\tan{\rho})} a_s (q_s, z,\omega_s), \nonumber \ \ \ \ \
\end{eqnarray}
where $\omega_{i} = \omega_{p} - \omega_{s} $, $q_{s,i}$ are the transverse components of the wavevectors, $k_{i, s, p} = n_{i,s,p} \omega_{i,s,p} / c$, $n_{i,s,p}$ are the refractive indices, $c$ is the speed of light,  $\Delta k_{\parallel} = \sqrt{k_{p}^2 - (q_s + q_i)^2} - \sqrt{k_s^2 - q_s^2} - \sqrt{k_i^2 - q_i^2}$ is the longitudinal phase matching, $\Delta k_{\perp} = q_s + q_i$ is the transverse phase matching, $\rho$ is the walk-off angle, and  $z \in [0, L]$.  Later, we neglect the frequency distribution of photons, 
assuming frequency filtering around $\omega_{p}/2$. The parameter $\Gamma$ is the coupling constant which depends on the nonlinear optical properties of the crystal and the intensity of the pump field, where only the latter can be varied in a wide range. The walk-off angle is defined as $\rho = \pm \arctan[ (n_{o}/n_{e})^2 \tan(\xi) ] \mp \xi $, where the upper signs refer to negative crystals ($n_{o}>n_{e}$) and the lower signs refer to the positive ones ($n_{e}>n_{o}$) with $n_{e}$ and $n_{o}$ being the extraordinary and the ordinary refractive indices of the crystal and $\xi $ denoting the angle between the pump wave vector and the optical axis of the crystal~\cite{Nikogosyan2005}.

In this paper, we consider BBO crystals with a ratio of the refractive indices  $n_{0}(\lambda_{i,s})/n_{e}(\lambda_{i,s}) = 1.0751$ and a  pump wavelength of $\lambda_{p} = 354.7$~nm. To close the phase matching synchronism for degenerate angles, the orientation of the crystal is set to  $\xi = 0.575$~rad which results in an anisotropy angle of $\rho = 0.067$~rad. The FWHM of the pump is taken to be 70~$\mu m$.

\begin{figure}[H]
 \includegraphics[width=0.85\textwidth,center]{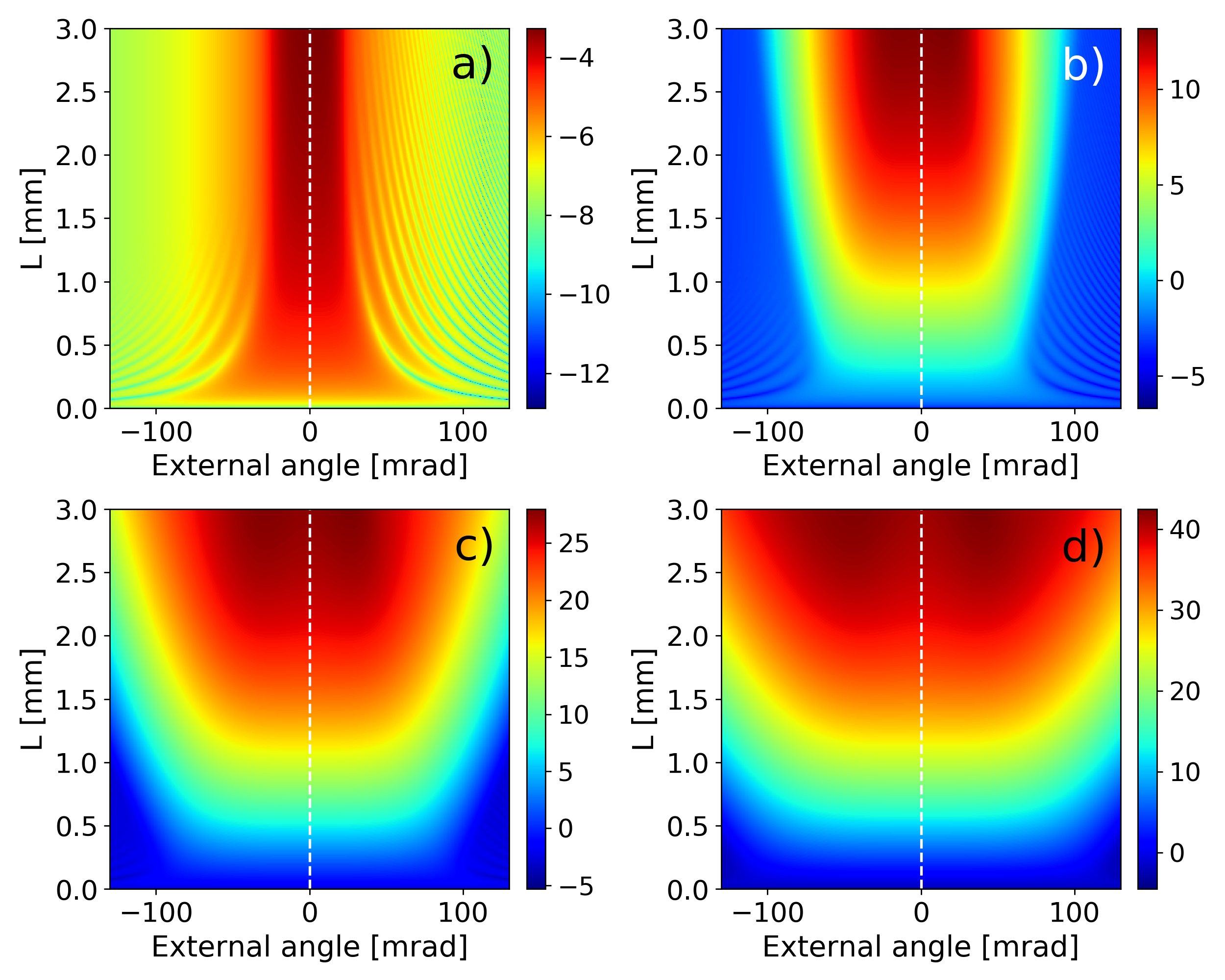}
  \caption{Angular intensity distributions on a logarithmic scale (base 10) as function of the crystal length $ L $. A single crystal with anisotropy is considered with gain parameters of a) $\Gamma$ = 0.001, b) $\Gamma$ = 0.2, c) $\Gamma$ = 0.4, and d) $\Gamma$ = 0.6.}
  \label{fig:intens_heatmap_z_angle}
\end{figure}

The solution to the system of integro-differential equations
Eqs.~(\ref{eqn:eq1_anis_1cr}) and (\ref{eqn:eq2_anis_1cr}) can be written in the following form:
\begin{eqnarray}
a_s(q_s, z, \omega_s) =
a_s(q_s) + \int \mathrm{d}q'_s \eta (q_s, q'_s, z,\Gamma) a_s(q'_s)+   \int \mathrm{d}q'_i \beta(q_s, q'_i, z,\Gamma) a^{\dagger}_i(q'_i),  \\
a^{\dagger}_i(q_i, z, \omega_i) =
a^{\dagger}_i(q_i) + \int \mathrm{d}q'_i \eta^{*} (q_i, q'_i, z,\Gamma) a^{\dagger}_i(q'_i) +   \int \mathrm{d}q'_s \beta^{*}(q_i, q'_s, z,\Gamma) a_s(q'_s), 
\label{solution1}
\end{eqnarray}
where $\eta(q_s, q'_s, z,\Gamma)$, $\beta(q_s, q'_i, z,\Gamma)$ are functions of the transverse wavevectors, the crystal length and the interaction strength $\Gamma$, while $a_s(q_s) = a_s(q_s, z=0, \omega_s)$ , $a^{\dagger}_i(q_i) = a^{\dagger}_i(q_i, z=0, \omega_i)$ are the initial plane-wave operators. For the case of an arbitrary pump and a high gain, the functions $\eta(q_s, q'_s, z,\Gamma)$ and $\beta(q_s, q'_i, z,\Gamma)$ can be found only numerically. Clearly, all observable quantities can be calculated by using this solution. For example, the angular intensity distribution of the signal beam is given by $  N_s(\theta_s)= \langle a^{\dagger}_s (\theta_s, z,\omega_s) a_s (\theta_s, z,\omega_s) \rangle = \int \mathrm{d}\theta_i' |\beta(\theta_s, \theta_i', z,\Gamma)|^2$, where $\theta_{s,i} = q_{s,i}/k_{s,i}$, and is shown in Fig.~\ref{fig:intens_heatmap_z_angle}  for different values of the parametric gain.

Inside the crystal, due to its birefringence, the extraordinary polarized pump beam is shifted,  see Fig.~\ref{fig:one_crystal}. At low gain, in the presence of anisotropy, an increase in the crystal length results in a stabilization of the angular intensity distribution: Its shape remains almost unchanged after a certain critical length of the crystal ($L_{cr} \approx 7$~mm for our parameters), see Fig.~\ref{fig:intens_heatmap_z_angle}a, while  the  width of the normalized angular intensity distribution $N_s (\theta)  / max(N_s(\theta) )$ quickly saturates, see Fig.~\ref{fig:1cr_fwhm_vs_gamma}. This arises since photons can be effectively amplified only in those regions where at least one of them overlaps with the pump beam, which, however, shifts in space. At low gain, due to phase matching synchronism, photons are emitted predominantly in the collinear direction. This means that after a certain distance, they leave the overlapping area and are no longer amplified. This fact strongly restricts the implementation of long crystals in the low intensity regime.

\section{Bright correlated twin-beam generation in a single crystal}

\begin{figure} [H]
  \includegraphics[width=0.6\textwidth,center]{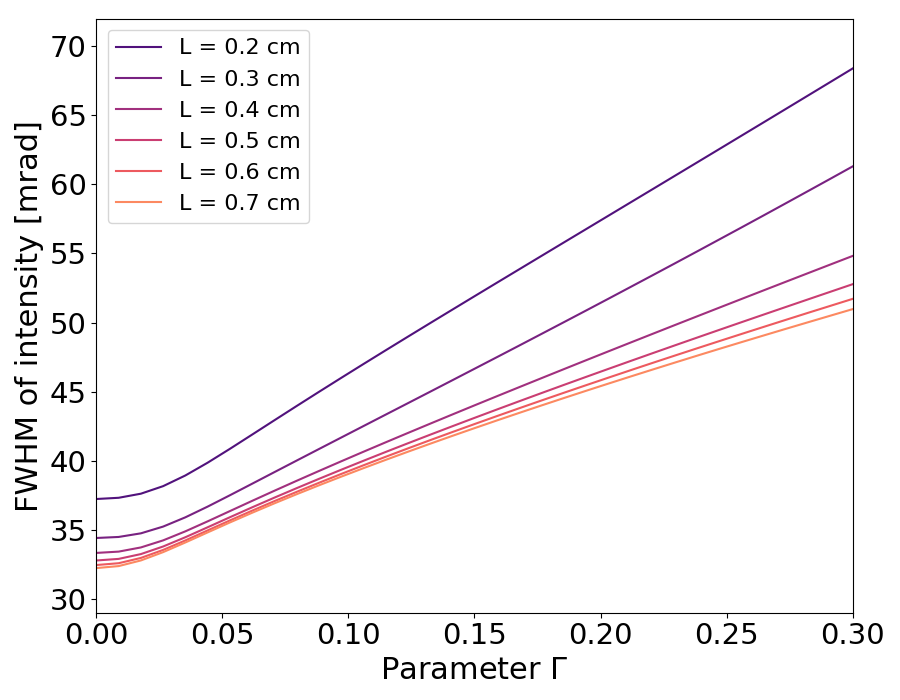}
  \caption{The FWHM of the angular intensity distribution as function of the gain parameter $\Gamma$ for  single crystals with anisotropy of  different lengths $L$.}
  \label{fig:1cr_fwhm_vs_gamma}
\end{figure}

However, an increase in the parametric gain dramatically changes the situation. With increasing gain, more noncollinear transverse wavevectors are involved in the interaction process, and the angular intensity distribution broadens, see Fig.~\ref{fig:intens_heatmap_z_angle}b-d. In the presence of walk-off, such broadening leads to a more efficient amplification of PDC photons with noncollinear angles of emission. The broadening strictly depends on the crystal length: the longer the crystal is, the more prominent the anisotropy effect becomes, the slower the intensity distribution broadens, see Fig.~\ref{fig:1cr_fwhm_vs_gamma}.

\begin{figure} [H]
  \includegraphics[width=1.05\textwidth,center]{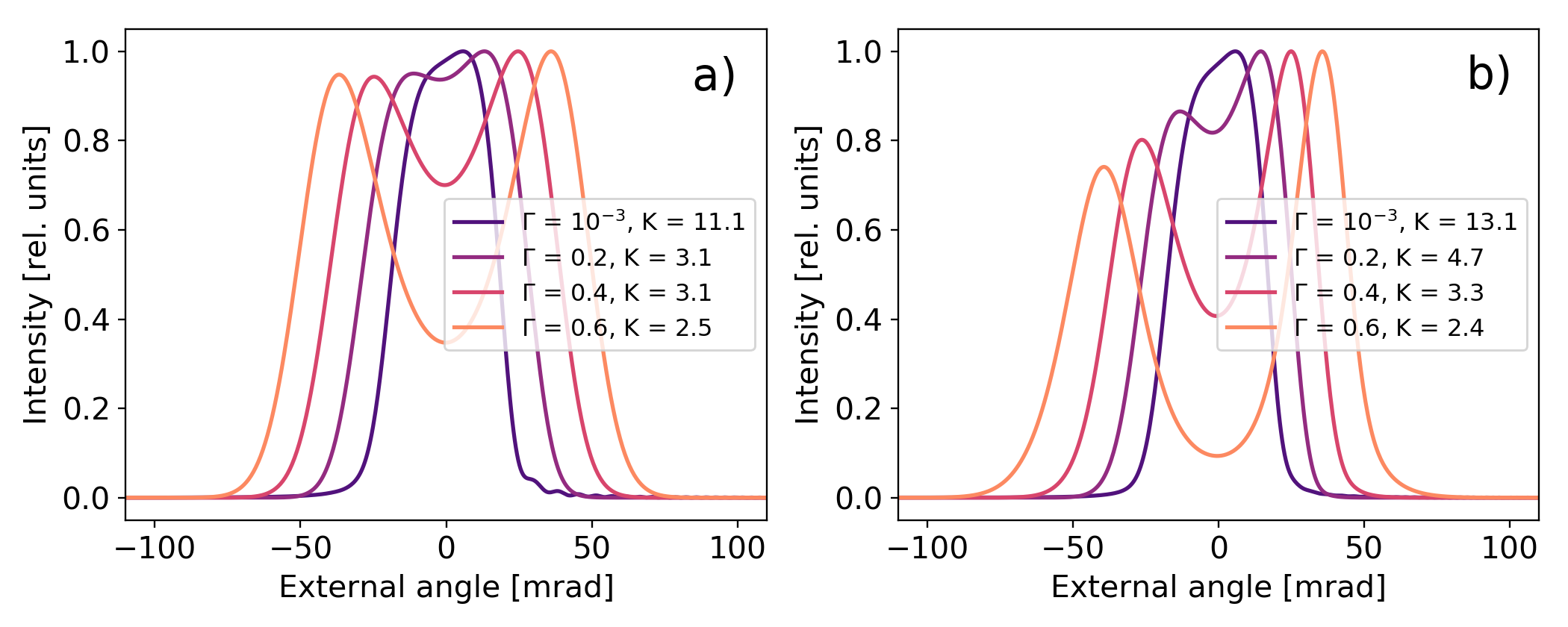}
  \caption{The angular intensity distributions for single crystals of lengths a) L = 0.2 cm and b) L = 0.3 cm for different gain parameters $\Gamma$. The longer the crystal is, the more pronounced the anisotropy effect becomes. The integral number of photons $\mathrm{N}_{\Sigma} = \int N(\theta) d\theta$ for gain parameters of $\Gamma$ = [0.001, 0.2, 0.4, 0.6] are
a) $\mathrm{N}_{\Sigma}$ = [$4.6 \times 10^{-5}$, $6 \times 10^{10}$, $1 \times 10^{24}$, $4.6 \times 10^{36}$] and b) $\mathrm{N}_{\Sigma}$ = [$7.4 \times 10^{-5}$, $1.7 \times 10^{12}$, $2.4 \times 10^{26}$, $6 \times 10^{39}$], respectively.}
  \label{fig:1cr_intens_diff_gains_with_anis}
\end{figure}

For a certain gain, the light intensity distribution is split into two beams, see Fig.~\ref{fig:1cr_intens_diff_gains_with_anis}. The first beam corresponds to the amplification in the walk-off direction, while the second one is generated in the symmetrical direction according to the phase matching synchronism. This splitting is a pure effect of the high intensity regime and cannot be observed at low gain, even in long crystals. As can be seen in Fig.~\ref{fig:1cr_intens_diff_gains_with_anis}, the splitting becomes more pronounced for higher gains. At the same time, the spatial asymmetry of generated PDC light (with respect to the collinear direction) progressively occurs with increasing crystal length since the signal photons are produced along the pump walk-off direction and thus amplified more strongly in comparison to the idler photons.

\subsection{Spatial mode structure}

The mode structure of the generated light can be understood by calculating the cross-correlation function between the signal and the idler beams: $\textnormal{cor}(\theta_{s}, \theta_{i}) = \langle N_{s}(\theta_{s})N_{i}(\theta_{i})\rangle - \langle N_{s}(\theta_{s})\rangle \langle N_{i}(\theta_{i})\rangle$.  The cross-correlation function $\textnormal{cor}(\theta_s, \theta_i)$ for an arbitrary gain can be decomposed into the Schmidt modes~\cite{PhysRevResearch.2.013371}:
\begin{equation} \label{eq:cov_decomposition}
\begin{aligned}
\textnormal{cor}(\theta_s, \theta_i, \Gamma) = \sum_{n} \sqrt{\Lambda_{n}} e^{i \phi_{n}} u_{n}(\theta_{s}, \Gamma) v_{n}(\theta_{i}, \Gamma),
\end{aligned}
\end{equation}
where $ u_{n}(\theta_{s}, \Gamma) $ and $ v_{n}(\theta_{i}, \Gamma) $ are the Schmidt modes of light for a fixed gain $\Gamma$, $\phi_{n}$ is the phase between $u_{n}$ and $v_{n}$, and $\Lambda_{n}(\Gamma)$ are the gain-dependent eigenvalues of the decomposition. The Schmidt number which quantifies the degree of spatial entanglement is defined through the gain-dependent eigenvalues as $ K = [\sum_{n} \Lambda_{n}^2]^{-1}$.

As one can observe in Fig.~\ref{fig:1cr_intens_diff_gains_with_anis}, a more pronounced anisotropy (longer crystals) leads to an increase in the number of Schmidt modes at low gain, which is why highly correlated beams cannot be obtained in the low-intensity regime by simply increasing the crystal length. Growth in the gain leads to the reduction in the number of modes, however, this reduction is faster for longer crystals. Finally, when the two bright peaks are spatially separated, the Schmidt number tends to $K=2$, indicating the generation of highly-correlated twin-beams.

Therefore, an increase in the parametric gain alone leads to the creation of bright highly-correlated twin-beams in long crystals with well pronounced anisotropy. However, to achieve a full splitting in a single crystal, quite high parametric gains are required, which in the experiment would destroy the nonlinear crystals. This situation might be improved by adding a small distance between two crystals, as will be shown in the next sections.

\section{Two-crystal system with anisotropy}

\begin{figure} [H]
  \includegraphics[width=0.8\textwidth,center]{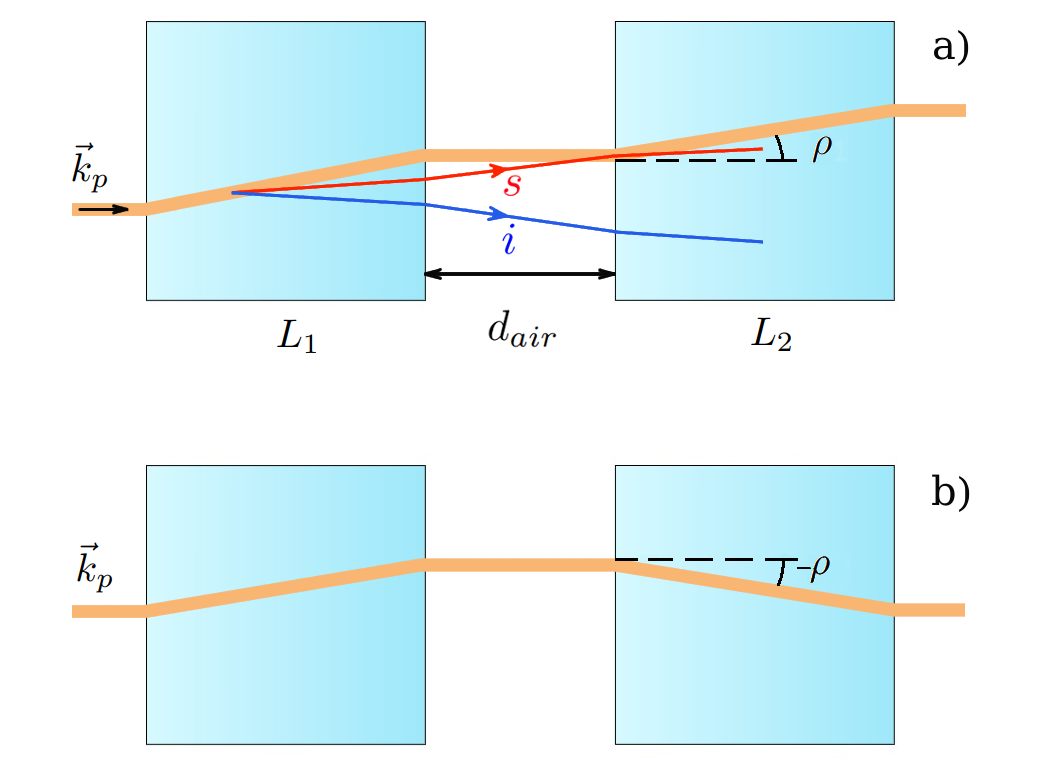}
  \caption{A schematic picture of two crystals of length $L_{1}$ and $L_{2}$ separated by the air gap of length $d$. a) Setup with non-compensated anisotropy. b) Setup with compensated anisotropy. The orange line shows the path of the pump beam, the red and the blue lines show paths of generated signal and idler photons, respectively.}
  \label{fig:2cr_scheme}
\end{figure}

For the case of  two crystals of length $L_{1}$ and $L_{2}$ with an in-between air gap of width $d$, see Fig.~\ref{fig:2cr_scheme}, the output operators in the high-gain regime could be found by considering the following steps. 1) Eqs.~(\ref{eqn:eq1_anis_1cr}) and (\ref{eqn:eq2_anis_1cr}),  should be solved for the first crystal with $z \in [0, L_{1}]$.  2) The solution from the first crystal should be taken as the initial condition for the second crystal. 3) The system of  equations similar to Eqs.~(\ref{eqn:eq1_anis_1cr}) and (\ref{eqn:eq2_anis_1cr}) should be solved for the second crystal with $z \in [0, L_{2}]$. Compared to Eqs.~(\ref{eqn:eq1_anis_1cr}) and (\ref{eqn:eq2_anis_1cr}), these equations include additional phases from the first crystal and the air gap.

For a system with non-compensated anisotropy, the systems of integro-differential equations for the second crystal is given by:
\begin{eqnarray}
\frac{\mathrm{d} a_s (q_s, z,\omega_s)}{\mathrm{d}z}=
\Gamma \int \mathrm{d}q_i \exp{-\frac{\sigma^2}{2}(\Delta k_{\parallel}\sin{\rho} + \Delta k_{\perp}\cos{\rho})^2 } \times \\ \exp{i \mu L_{1}    + i\Delta k^{air} d + i \mu z} a^{\dagger}_i(q_i, z,\omega_i), \nonumber \ \ \ \ \
\label{eq1_anis_noncomp_2nd_cr}
\end{eqnarray}
\begin{eqnarray}
\frac{\mathrm{d} a^{\dagger}_i(q_i, z,\omega_i)}{\mathrm{d}z}=
\Gamma \int \mathrm{d}q_s \exp{-\frac{\sigma^2}{2}( \Delta k_{\parallel}\sin{\rho} + \Delta k_{\perp}\cos{\rho})^2 } \times \\ \exp{-i \mu L_{1}  -i\Delta k^{air} d - i \mu z  } a_s (q_s, z,\omega_s), \nonumber \ \ \ \ \
\label{eq2_anis_noncomp_2nd_cr}
\end{eqnarray}
where $z \in [0, L_{2}]$,  $\Delta k^{air} = \sqrt{(k_{p}^{air})^2 - (q_s^{air} + q_i^{air} )^2} - \sqrt{(k_s^{air})^2 - (q_s^{air} )^2 } - \sqrt{(k_i^{air})^2  - (q_i^{air} )^2}$ is the phase matching in the air,  $d$ is the distance between the crystals, $\mu = \Delta k_{\parallel} - \Delta k_{\perp}\tan{\rho}$. For the system with compensated anisotropy, the walk-off angle $\rho$ in the Gaussian exponents has to be changed to $-\rho$, while $\mu$ in the last term of the imaginary exponents has to be changed to $\xi=\Delta k_{\parallel} + \Delta k_{\perp}\tan{\rho}$. The explicit expressions for the equations in the second crystal for the case of a compensated anisotropy are presented in the Appendix.

\section{Bright correlated twin-beam generation in a two-crystal system}

\begin{figure}
  \includegraphics[width=1.1\textwidth,center]{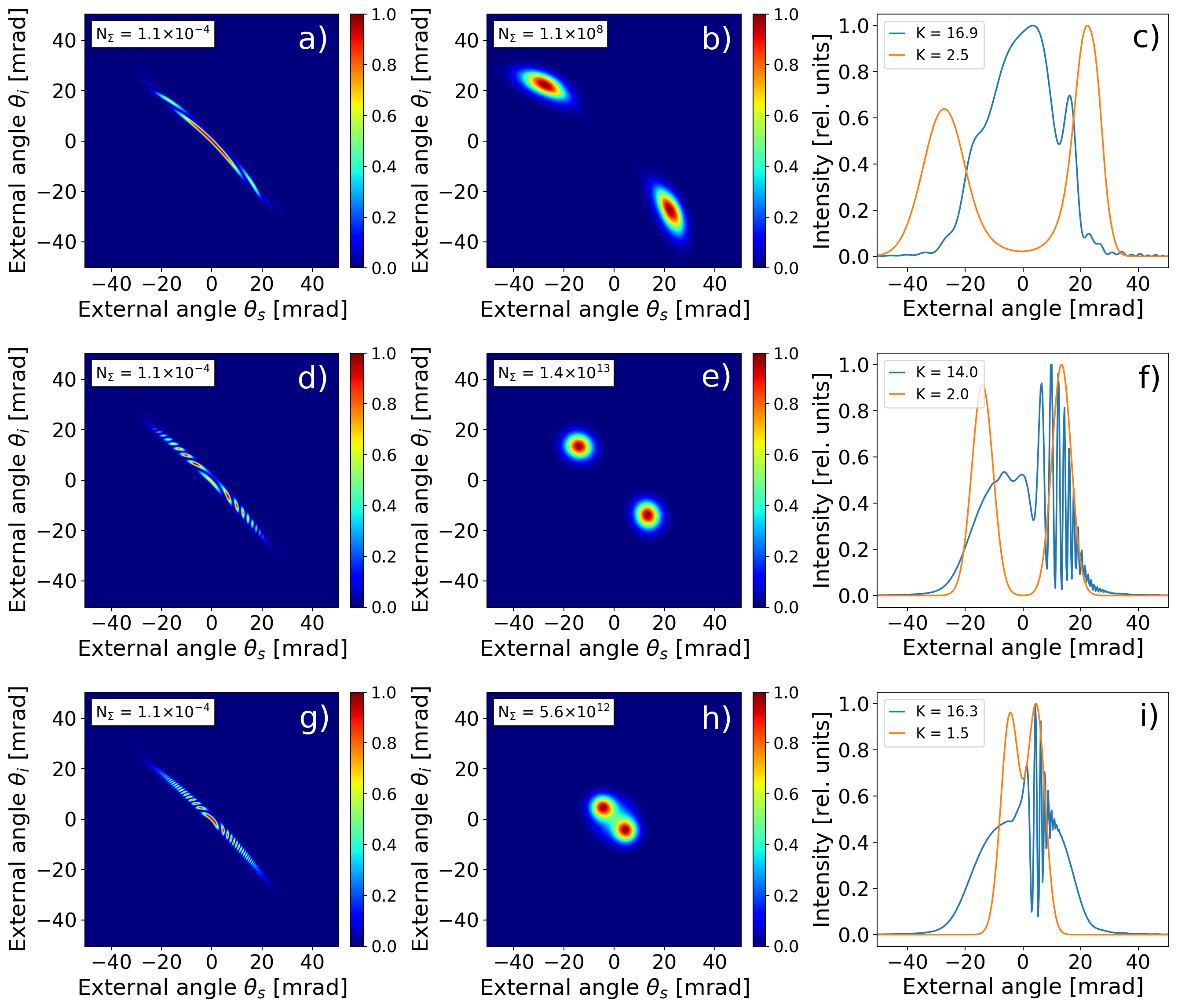}
  \caption{The calculated correlation functions at low  ($\Gamma = 0.001$, left column) and high   ($\Gamma = 0.12$, middle column) gain, and the corresponding intensity distributions (right column). The non-compensated scheme with $L_{1} = L_{2} = 0.25$~cm is considered. The distance between the crystals is a-c) d = $0.15$~cm, d-f) d = $1.11$~cm, and g-i) d = $3.51$~cm.}
  \label{fig:2corr_1intens_l-0.25_d-0.15}
\end{figure}

In section 2, we discussed how two-mode correlated beams could be generated in a single crystal by increasing the gain. Interestingly, adding a small distance between two crystals (the non-compensated configuration) strongly reduces the required pump intensities due to the overlap of noncollinearly emitted PDC photons with the pump in the second crystal. Indeed, as it is shown in Fig.~\ref{fig:2cr_scheme}a, noncollinear photons in a single crystal quickly leave the overlapping region. However, the air gap allows such photons to overlap again with the pump in the second crystal, resulting in amplification of light in this noncollinear direction as well as in the symmetrical one (according to the phase matching synchronism). Such amplification of noncollinear light produces separated correlated beams, see Fig.~\ref{fig:2corr_1intens_l-0.25_d-0.15}, which is only possible in the high-gain regime where strongly noncollinear photons are involved in the interaction process. 

\begin{figure} [H]
  \includegraphics[width=1.0\textwidth,center]{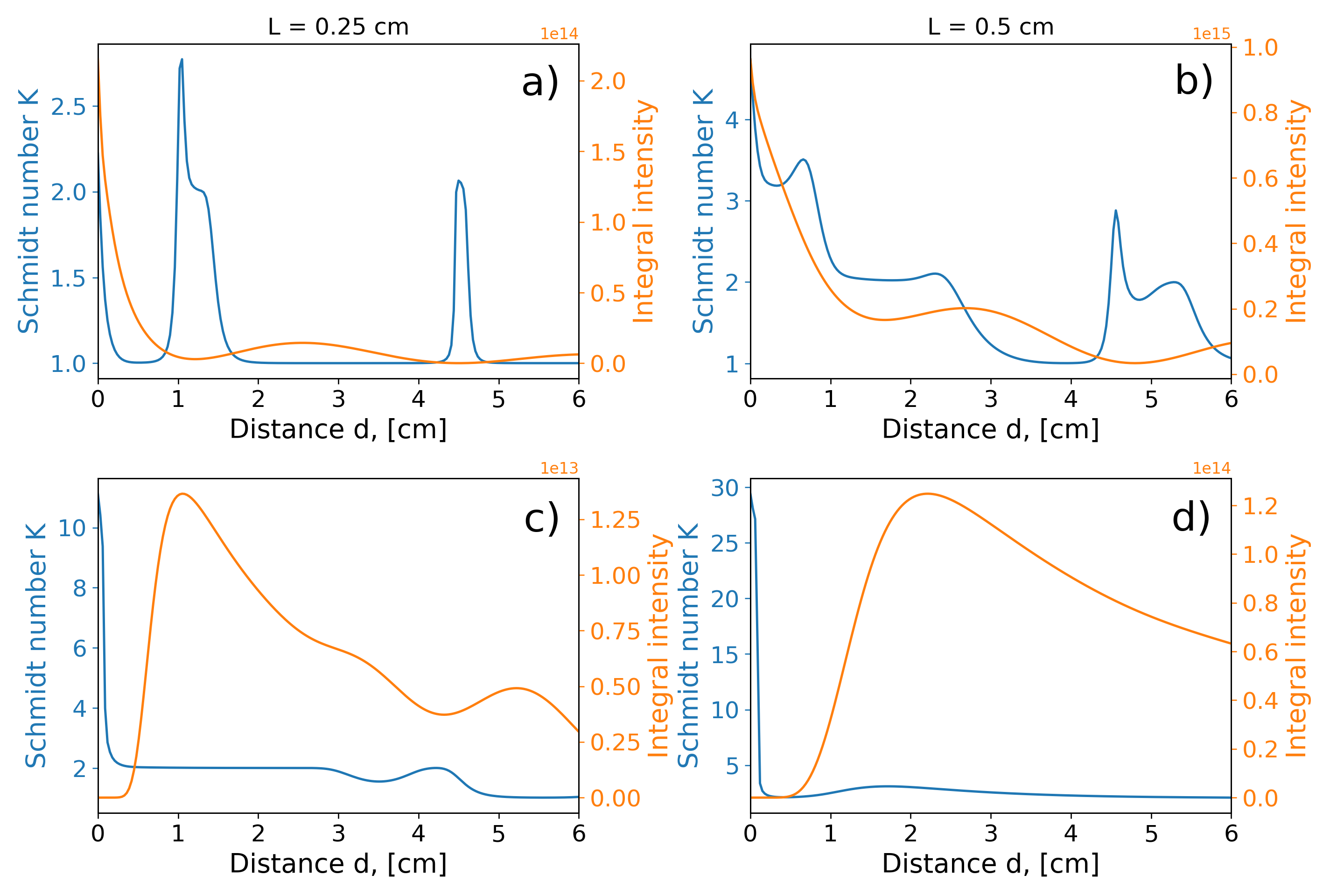}
  \caption{The calculated Schmidt number (blue) and the integral intensity (orange) as functions of the distance $d$ for  $L_{1} = L_{2} = 0.25$~cm (a, c) and $L_{1} = L_{2} = 0.5$~cm (b, d). The compensated setup is presented in a and b, while the   non-compensated setup is used in c and d. The gain parameter is $\Gamma$ = 0.12.}
  \label{fig:shm_num_vs_d_l-0.25_0.5}
\end{figure}

To characterize the mode structure of generated light, we calculate the Schmidt number through the decomposition of the cross-correlation function Eq.~(\ref{eq:cov_decomposition}). For the non-compensated configuration, the cross-correlation functions in the low- and high-gain regimes are presented in the left and the middle columns of  Fig.~\ref{fig:2corr_1intens_l-0.25_d-0.15}, respectively. One can observe that for the chosen pump bandwidth FWHM = 70~$\mu$m, the radiation is highly multimode at low gain, regardless of the distance between the crystals. However, with increasing gain, the number of modes dramatically reduces. For example, for $d=0.15$~cm, the correlation function at high gain has two peaks that are separated from each other, which results in the Schmidt number $K=2.5$, see Fig.~\ref{fig:2corr_1intens_l-0.25_d-0.15}b. In such a case, the intensity distribution profile also has a two-beam shape, which is completely different from the case without a distance, where the intensity distribution has an asymmetric single-peak shape, Fig.~\ref{fig:1cr_intens_diff_gains_with_anis}.  This means that contrary to the low intensity regime, placing a small distance between the two crystals at high gain strongly modifies the structure of the generated light. Indeed, as can be seen in Fig.~\ref{fig:shm_num_vs_d_l-0.25_0.5}c,d, starting from $d \approx 0.15$~cm, the Schmidt number drops down and with further increase in distance approaches $K=2$. The higher the gain is, the more separated the intensity peaks are for a fixed distance $d$, and the shorter the distance $d$ becomes, which is required to achieve two modes, as photons with larger angles of emission are involved in the interaction process.

For a fixed gain, an increase in the distance between crystals reduces the number of modes and leads to the generation of perfectly correlated distinct peaks, see Fig.~\ref{fig:2corr_1intens_l-0.25_d-0.15} e,f. At the same time, there is a strong rise in the integral intensity with growing distance, see Fig.~\ref{fig:shm_num_vs_d_l-0.25_0.5}. Indeed, in the presence of anisotropy, due to the walk-off effect, a single crystal has a saturation intensity limit: There is a critical crystal length, after which the intensity no longer rises with increasing crystal length. A distance between crystals allows noncollinear photons to overlap again with the pump in the second crystal, which leads to an increase in the integral intensity for the non-compensated configuration compared to a single crystal of double length.  As one can observe in Fig.~\ref{fig:shm_num_vs_d_l-0.25_0.5}c, adding a distance results in an amplification of the integral intensity from $N_{\Sigma} = 10^{9}$ for $d=0$~cm up to $N_{\Sigma} = 10^{13}$ for $d=1.1$~cm. The maximum integral intensity is obtained for the distance $d'$ between the crystals that provides the best overlap between the pump and PDC light in the second crystal. This distance is connected with the amplified external angle $\theta'$ as $ \theta' L /n_p + \theta' d' = \rho L$. For $\Gamma=0.12$ the amplified external  angle is about $\theta'=0.014$~rad, Fig.~\ref{fig:2corr_1intens_l-0.25_d-0.15}f, which results in $d'=1.1$~cm for $L=0.25$~cm crystals.  At the same time, a twofold increase in the crystal length leads to a doubling of the distance $ d' $, namely, to $ d' = 2.2 $~cm, as  shown in Fig.~\ref{fig:shm_num_vs_d_l-0.25_0.5}c,d.

A further increase in the distance for the non-compensated configuration leads to stronger scattering of noncollinear photons from the first crystal. This is why only near collinear photons from the first crystal overlap with the pump in the second crystal, which results in a reduction of the number of modes and a tendency towards the single-mode regime. For example, for $d=3.51$~cm the high-gain regime leads to $K=1.5$, see Fig.~\ref{fig:2corr_1intens_l-0.25_d-0.15}h,i. Starting from distances $d \approx 5$~cm, see Fig.~\ref{fig:shm_num_vs_d_l-0.25_0.5}c, a single-mode regime ($K=1$) is achieved. Longer crystals tend to generate narrower modes.

When the second crystal is placed for compensation, the mode structure of generated light changes, the compensated configuration pursues the goal of collinear emission amplification, leading to a single-mode generation at high gain. This is why the Schmidt number for high gain drops to $K=1$ by adding a distance, and only in special cases, when two degenerate modes are created ($d \approx 1.1$~cm, $d \approx 4.5$~cm and further), the Schmidt number reaches $K=2$, see Fig.~\ref{fig:shm_num_vs_d_l-0.25_0.5}a. These special distances lead to a phase difference between the pump and PDC light that suppresses the emission in the collinear direction after the second crystal and results in two degenerate twin-beams for short crystals. However, as the length of the crystal increases, the structure of the modes changes since the compensation no longer works properly, see Fig.~\ref{fig:shm_num_vs_d_l-0.25_0.5}b. Less efficient compensation results in an increase of the number of modes: Only large distances $d$ lead to the single-mode regime, since the part of the radiation from the first crystal that overlaps with the pump beam in the second crystal is getting cut smaller due to scattering. For very long distances, this part has a plane wavefront.

\section{Comparison between the compensated and non-compensated configurations}

\begin{figure}
  \includegraphics[width=1.05\textwidth,center]{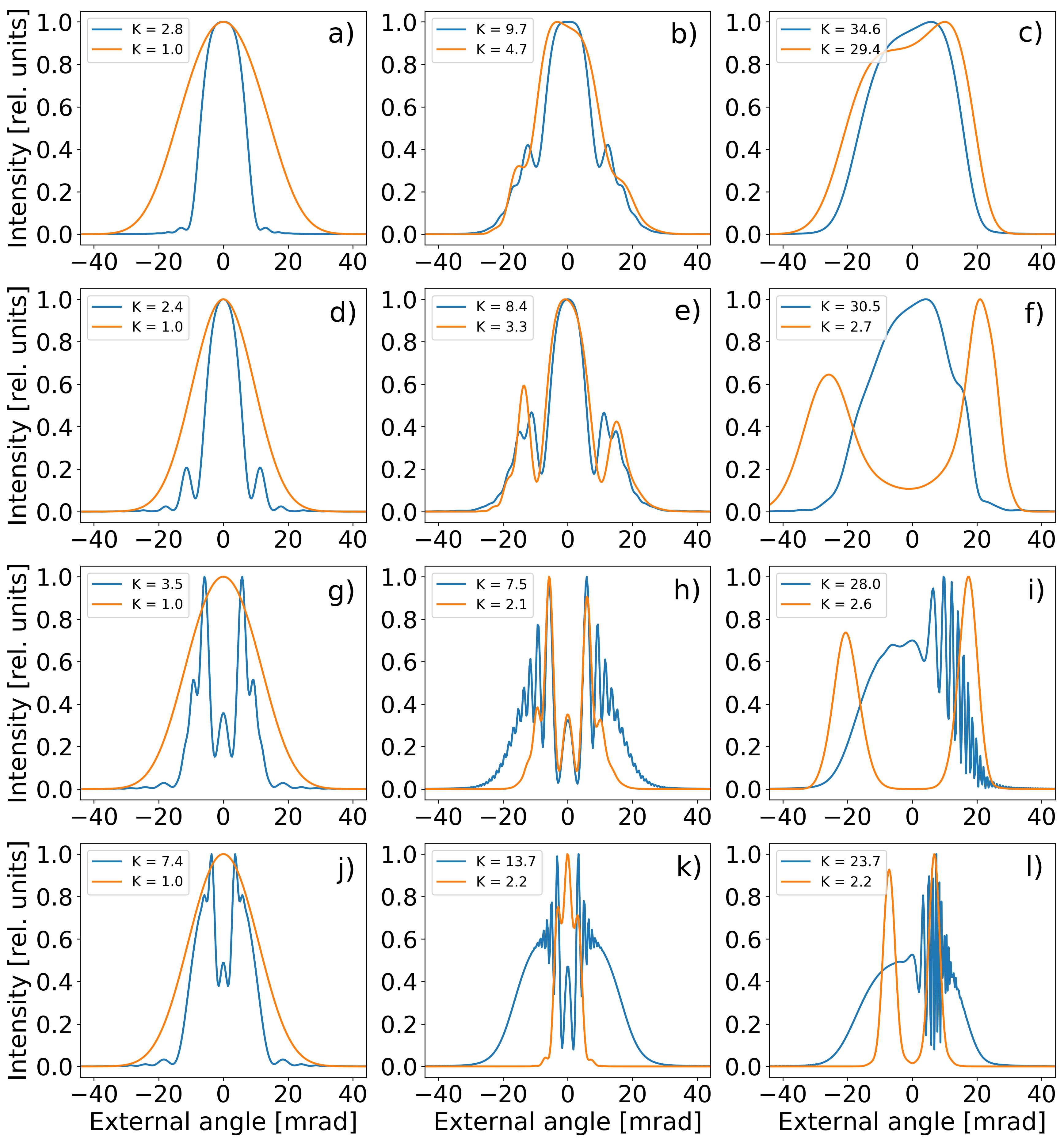}
  \caption{The calculated intensity distributions for two crystals of length L = 0.5 cm and an air gap of length a-c) d = 0.0~cm, d-f) d = 0.15~cm, and g-i) d = 1.11~cm, and j-l) 4.5~cm. Blue curves correspond to the low gain, $\Gamma = 0.001$, while orange curves describe the high gain, $\Gamma = 0.12$. The left column shows the case where anisotropy is neglected, the middle column presents the anisotropy-compensated configuration, while the right column corresponds to the anisotropy non-compensated configuration. The Schmidt numbers for each case are shown in the legend.}
  \label{fig:3cases_lg_hg_l-0.5_d-0.0}
\end{figure}

In this section, we discuss the intensity distributions of light generated in the compensated and non-compensated configurations and also demonstrate the importance of taking anisotropy into account for long crystals by comparing the results to a model where the anisotropy is neglected. The anisotropy-compensated configuration results in symmetrical intensity distributions in the low-gain regime, see the middle column of Fig.~\ref{fig:3cases_lg_hg_l-0.5_d-0.0}. However, for higher gains, the compensation stops working as most of the intensity is generated in the second crystal. This is reflected in the asymmetry of the spectrum for high gains, see orange curves in the middle column of Fig.~\ref{fig:3cases_lg_hg_l-0.5_d-0.0}. As is well-known, the model without anisotropy fails to describe long crystals properly. As can be seen in the left and middle columns of Fig.~\ref{fig:3cases_lg_hg_l-0.5_d-0.0}, this deviation becomes more pronounced with increasing the distance between crystals, as well as at high gains. The non-compensated anisotropy leads to asymmetry in the intensity distribution. However, with growing distance, this asymmetry decreases.

The distance between crystals leads to an additional phase which results in interference for both the compensated and the non-compensated configurations. However, increasing the gain affects the two schemes differently: While in the compensated configuration, a high gain tends to narrow the intensity distribution, in the non-compensated configuration, a high gain results in peak separation.
For the fixed pump profile, there are three main factors that define the angular shape of the generated light: The length of the crystals, the distance between them, and the parametric gain. Generally, an increase in the gain broadens the intensity distribution, while an increase in the distance narrows it. This is why there exist a distance $d \approx 0.15 $~cm at which these two opposite effects basically eliminate each other in the compensated configuration, see Fig.~\ref{fig:3cases_lg_hg_l-0.5_d-0.0}e. One can see that an increase of the gain does almost not change the shape of the intensity in this case. In the non-compensated configuration, at a distance of about $d \approx 0.15$~cm, the splitting of beams begins in the high-gain regime. 

The distance $d=1.11$~cm already gives two fully separated beams with $K=2$ for the non-compensated configuration. The compensated configuration with $d=1.11$~cm also results in a two-mode light, see Fig.~\ref{fig:shm_num_vs_d_l-0.25_0.5}b, which, however, does not correspond to the separated correlated beams in the case of long crystals. A further increase in the distance leads to a loss of amplification of noncollinear angles in the second crystal at high gain: The intensity distribution in both compensated and non-compensated configurations becomes narrower, see Fig.~\ref{fig:3cases_lg_hg_l-0.5_d-0.0}k,l, and finally tends to the single-mode regime, as shown in Fig.~\ref{fig:shm_num_vs_d_l-0.25_0.5}. However, in the low-gain regime, the generated light is strongly multimode even for large distances, which is reflected in the broad intensity distributions with fast interference fringes.

\section*{Conclusion}

We investigate properties of bright squeezed light generated in the high-gain PDC process within long crystals with strong anisotropy. A detailed theoretical analysis of the generated light based on systems of integro-differential equations for the plane-wave operators, which take into account the pump walk-off inside nonlinear crystals, is presented. We demonstrate that at high gain, the undesirable anisotropy effect can be used in a positive way to generate highly-correlated bright twin-beams. Such twin-beams already arise in a single crystal. However, placing a small distance between the crystals allows one to generate correlated twin-beams at already rather low intensities of the pump. At the same time, by adding such a distance, one can significantly (by several orders of magnitude) increase the intensity of twin-beams in comparison with systems without an air gap. Finally, the shape of the intensity distributions and the mode structure of the generated light were investigated and analyzed in systems with compensated and non-compensated anisotropy for various gains, air gaps, and crystal lengths.

\section*{Acknowledgments}

Financial support by the Deutsche Forschungsgemeinschaft (DFG) through the TRR 142/2 (project number 231447078, project C06) and the project SH 1228/3-1 is gratefully acknowledged.
We also thank the PC$^2$ (Paderborn Center for Parallel Computing) for providing computing time.

\normalem
\bibliographystyle{unsrt}
\bibliography{bibliography}

\begin{thebibliography}{10}
\newcommand{\enquote}[1]{``#1''}

\bibitem{PhysRevLett.25.84}
D.~C. Burnham and D.~L. Weinberg, \enquote{Observation of simultaneity in
  parametric production of optical photon pairs,} {\protect\JournalTitle{Phys.
  Rev. Lett.}} \textbf{25}, 84--87 (1970).

\bibitem{PhysRev.124.1646}
W.~H. Louisell, A.~Yariv, and A.~E. Siegman, \enquote{Quantum fluctuations and
  noise in parametric processes. {I}.} {\protect\JournalTitle{Phys. Rev.}}
  \textbf{124}, 1646--1654 (1961).

\bibitem{Karan_2020}
S.~Karan, S.~Aarav, H.~Bharadhwaj, L.~Taneja, A.~De, G.~Kulkarni, N.~Meher, and
  A.~K. Jha, \enquote{Phase matching in $\upbeta$-barium borate crystals for
  spontaneous parametric down-conversion,} {\protect\JournalTitle{Journal of
  Optics}} \textbf{22}, 083501 (2020).

\bibitem{WALBORN201087}
S.~Walborn, C.~Monken, S.~Pádua, and P.~{Souto Ribeiro}, \enquote{Spatial
  correlations in parametric down-conversion,} {\protect\JournalTitle{Physics
  Reports}} \textbf{495}, 87 -- 139 (2010).

\bibitem{PhysRevA.91.043816}
P.~Sharapova, A.~M. P\'erez, O.~V. Tikhonova, and M.~V. Chekhova,
  \enquote{Schmidt modes in the angular spectrum of bright squeezed vacuum,}
  {\protect\JournalTitle{Phys. Rev. A}} \textbf{91}, 043816 (2015).

\bibitem{PhysRevA.97.053827}
P.~R. Sharapova, O.~V. Tikhonova, S.~Lemieux, R.~W. Boyd, and M.~V. Chekhova,
  \enquote{Bright squeezed vacuum in a nonlinear interferometer: Frequency and
  temporal schmidt-mode description,} {\protect\JournalTitle{Phys. Rev. A}}
  \textbf{97}, 053827 (2018).

\bibitem{PhysRevA.69.023802}
E.~Brambilla, A.~Gatti, M.~Bache, and L.~A. Lugiato, \enquote{Simultaneous
  near-field and far-field spatial quantum correlations in the high-gain regime
  of parametric down-conversion,} {\protect\JournalTitle{Phys. Rev. A}}
  \textbf{69}, 023802 (2004).

\bibitem{PhysRevLett.93.243601}
O.~Jedrkiewicz, Y.-K. Jiang, E.~Brambilla, A.~Gatti, M.~Bache, L.~A. Lugiato,
  and P.~Di~Trapani, \enquote{Detection of sub-shot-noise spatial correlation
  in high-gain parametric down conversion,} {\protect\JournalTitle{Phys. Rev.
  Lett.}} \textbf{93}, 243601 (2004).

\bibitem{PhysRevLett.99.063901}
M.~V. Fedorov, M.~A. Efremov, P.~A. Volkov, E.~V. Moreva, S.~S. Straupe, and
  S.~P. Kulik, \enquote{Anisotropically and high entanglement of biphoton
  states generated in spontaneous parametric down-conversion,}
  {\protect\JournalTitle{Phys. Rev. Lett.}} \textbf{99}, 063901 (2007).

\bibitem{Huo_and_Zhang}
G.~Huo, T.~Zhang, M.~Zhang, H.~Zhu, X.~Zha, and H.~Chen, \enquote{Anisotropic
  narrowing of biphoton wave packet distribution in spontaneous parametric
  down-conversion with biaxial crystal,} {\protect\JournalTitle{Journal of
  Modern Optics}} \textbf{65}, 298--305 (2018).

\bibitem{PhysRevA.77.015805}
G.~Brida, M.~Genovese, M.~V. Chekhova, and L.~A. Krivitsky, \enquote{Tailoring
  polarization entanglement in anisotropy-compensated spontaneous parametric
  down-conversion,} {\protect\JournalTitle{Phys. Rev. A}} \textbf{77}, 015805
  (2008).

\bibitem{P_rez_2013}
A.~M. P{\'{e}}rez, F.~Just, A.~Cavanna, M.~V. Chekhova, and G.~Leuchs,
  \enquote{Compensation of anisotropy effects in a nonlinear crystal for
  squeezed vacuum generation,} {\protect\JournalTitle{Laser Physics Letters}}
  \textbf{10}, 125201 (2013).

\bibitem{Cavanna:14}
A.~Cavanna, A.~M. P\'{e}rez, F.~Just, M.~V. Chekhova, and G.~Leuchs,
  \enquote{Compensation of anisotropy effects in the generation of two-photon
  light,} {\protect\JournalTitle{Opt. Express}} \textbf{22}, 9983--9992 (2014).

\bibitem{Perez2015}
A.~M. P{\'e}rez, K.~Y. Spasibko, P.~R. Sharapova, O.~V. Tikhonova, G.~Leuchs,
  and M.~V. Chekhova, \enquote{Giant narrowband twin-beam generation along the
  pump-energy propagation direction,} {\protect\JournalTitle{Nature
  Communications}} \textbf{6}, 7707 (2015).

\bibitem{Hochrainer1508}
A.~Hochrainer, M.~Lahiri, R.~Lapkiewicz, G.~B. Lemos, and A.~Zeilinger,
  \enquote{Quantifying the momentum correlation between two light beams by
  detecting one,} {\protect\JournalTitle{Proceedings of the National Academy of
  Sciences}} \textbf{114}, 1508--1511 (2017).

\bibitem{Rarity1992}
J.~G. Rarity, P.~R. Tapster, J.~A. Levenson, J.~C. Garreau, I.~Abram, J.~Mertz,
  T.~Debuisschert, A.~Heidmann, C.~Fabre, and E.~Giacobino, \enquote{Quantum
  correlated twin beams,} {\protect\JournalTitle{Applied Physics B}}
  \textbf{55}, 250--257 (1992).

\bibitem{Pradyumna2020}
S.~T. Pradyumna, E.~Losero, I.~Ruo-Berchera, P.~Traina, M.~Zucco, C.~S.
  Jacobsen, U.~L. Andersen, I.~P. Degiovanni, M.~Genovese, and T.~Gehring,
  \enquote{Twin beam quantum-enhanced correlated interferometry for testing
  fundamental physics,} {\protect\JournalTitle{Communications Physics}}
  \textbf{3}, 104 (2020).

\bibitem{Clark_2021}
A.~S. Clark, M.~Chekhova, J.~C.~F. Matthews, J.~G. Rarity, and R.~F. Oulton,
  \enquote{Special topic: Quantum sensing with correlated light sources,}
  {\protect\JournalTitle{Applied Physics Letters}} \textbf{118}, 060401 (2021).

\bibitem{Jason_D_Muellera}
J.~D. Muellera, N.~Samantaray, and J.~C.~F. Matthews, \enquote{A practical
  model of twin-beam experiments for sub-shot-noise absorption measurements,}
  {\protect\JournalTitle{Applied Physics Letters}} \textbf{117}, 034001 (2020).

\bibitem{Brida2010}
G.~Brida, M.~Genovese, and I.~Ruo~Berchera, \enquote{Experimental realization
  of sub-shot-noise quantum imaging,} {\protect\JournalTitle{Nature Photonics}}
  \textbf{4}, 227--230 (2010).

\bibitem{Celebrano2011}
M.~Celebrano, P.~Kukura, A.~Renn, and V.~Sandoghdar, \enquote{Single-molecule
  imaging by optical absorption,} {\protect\JournalTitle{Nature Photonics}}
  \textbf{5}, 95--98 (2011).

\bibitem{Whittaker_2017}
R.~Whittaker, C.~Erven, A.~Neville, M.~Berry, J.~L. O'Brien, H.~Cable, and
  J.~C.~F. Matthews, \enquote{Absorption spectroscopy at the ultimate quantum
  limit from single-photon states,} {\protect\JournalTitle{New Journal of
  Physics}} \textbf{19}, 023013 (2017).

\bibitem{Losero2018}
E.~Losero, I.~Ruo-Berchera, A.~Meda, A.~Avella, and M.~Genovese,
  \enquote{Unbiased estimation of an optical loss at the ultimate quantum limit
  with twin-beams,} {\protect\JournalTitle{Scientific Reports}} \textbf{8},
  7431 (2018).

\bibitem{Agliati_2005}
A.~Agliati, M.~Bondani, A.~Andreoni, G.~D. Cillis, and M.~G.~A. Paris,
  \enquote{Quantum and classical correlations of intense beams of light
  investigated via joint photodetection,} {\protect\JournalTitle{Journal of
  Optics B: Quantum and Semiclassical Optics}} \textbf{7}, S652--S663 (2005).

\bibitem{Li_20}
F.~Li, T.~Li, and G.~S. Agarwal, \enquote{Temporal quantum noise reduction
  acquired by an electron-multiplying charge-coupled-device camera,}
  {\protect\JournalTitle{Opt. Express}} \textbf{28}, 37538--37545 (2020).

\bibitem{Chekhova_18}
M.~V. Chekhova, S.~Germanskiy, D.~B. Horoshko, G.~K. Kitaeva, M.~I. Kolobov,
  G.~Leuchs, C.~R. Phillips, and P.~A. Prudkovskii, \enquote{Broadband bright
  twin beams and their upconversion,} {\protect\JournalTitle{Opt. Lett.}}
  \textbf{43}, 375--378 (2018).

\bibitem{Wu_19}
M.-C. Wu, B.~L. Schmittberger, N.~R. Brewer, R.~W. Speirs, K.~M. Jones, and
  P.~D. Lett, \enquote{Twin-beam intensity-difference squeezing below 10 {H}z,}
  {\protect\JournalTitle{Opt. Express}} \textbf{27}, 4769--4780 (2019).

\bibitem{PhysRevResearch.2.013371}
P.~R. Sharapova, G.~Frascella, M.~Riabinin, A.~M. P\'erez, O.~V. Tikhonova,
  S.~Lemieux, R.~W. Boyd, G.~Leuchs, and M.~V. Chekhova, \enquote{Properties of
  bright squeezed vacuum at increasing brightness,}
  {\protect\JournalTitle{Phys. Rev. Research}} \textbf{2}, 013371 (2020).

\bibitem{PhysRevA.81.053805}
R.~S. Bennink, \enquote{Optimal collinear {G}aussian beams for spontaneous
  parametric down-conversion,} {\protect\JournalTitle{Phys. Rev. A}}
  \textbf{81}, 053805 (2010).

\bibitem{Nikogosyan2005}
D.~N. Nikogosyan, \emph{Nonlinear Optical Crystals: A Complete Survey}
  (Springer, New York, 2005).

\bibitem{Loudon:105699}
R.~Loudon, \emph{{The Quantum Theory of Light}} (Clarendon Press, Oxford,
  1973).

\end{thebibliography}

\section*{Appendix}

\subsection*{A single crystal with anisotropy at low gain. The two-photon amplitude approach.}

At low parametric gain, the wave-function of light generated in the PDC process can be described by its two-photon amplitude $F(\textbf{\textit{k}}_{s}, \textbf{\textit{k}}_{i}) $~\cite{Loudon:105699}.

\begin{equation} \label{eq:wave_func}
\begin{aligned}
& | \psi \rangle \propto \int d\textbf{\textit{k}}_{s} d\textbf{\textit{k}}_{i} F(\textbf{\textit{k}}_{s}, \textbf{\textit{k}}_{i}) \hat{a}^{\dag}(\textbf{\textit{k}}_{s}) \hat{a}^{\dag}(\textbf{\textit{k}}_{i}) |0 \rangle. \\
\end{aligned}
\end{equation}

For a single crystal of length $L$, the two-photon amplitude that takes into account anisotropy can be found analytically~\cite{P_rez_2013}:

\begin{equation} \label{eq:ampl_single_crystal}
\begin{aligned}
& F(\textbf{\textit{k}}_{s}, \textbf{\textit{k}}_{i}) \propto \exp[-\frac{\sigma_{p}^2}{2}(\Delta k_{\parallel}\sin{\rho} + \Delta k_{\perp}\cos{\rho})^2] \textrm{sinc}(\frac{L}{2}(\Delta k_{\parallel} - \Delta k_{\perp}\tan{\rho})), \\
\end{aligned}
\end{equation}
where $\Delta k_{\parallel} = \sqrt{k_{p}^2 - (q_s + q_i)^2} - \sqrt{k_s^2 - q_s^2} - \sqrt{k_i^2 - q_i^2}$, $k_p = n_p  \omega_p / c, ~
k_{i, s} = n_{i,s} \omega_{i,s} / c$,  $\Delta k_{\perp} = q_s + q_i$ and $q_{s,i}$ are the  transverse components of the wavevectors. 

\subsection*{A two-crystal system with an air gap and anisotropy at low gain}

For two crystals with an air gap of length $d$ in-between, see Fig.~\ref{fig:2cr_scheme}a, the biphoton amplitude at low gain can be written as a sum of amplitudes of individual crystals with the phase difference acquired in the air gap:
\begin{equation} \label{eq:ampl_two_crystals}
\begin{aligned}
& F(\textbf{\textit{k}}_{s}, \textbf{\textit{k}}_{i}) \propto F_{1}(\textbf{\textit{k}}_{s}, \textbf{\textit{k}}_{i}) + \exp(i\Delta k^{air} d) F_{2}(\textbf{\textit{k}}_{s}, \textbf{\textit{k}}_{i}),  \\
\end{aligned}
\end{equation}
where $\Delta k^{air} = \sqrt{(k_{p}^{air})^2 - (q_s^{air} + q_i^{air})^2} - \sqrt{(k_s^{air})^2 - (q_s^{air})^2 } - \sqrt{(k_i^{air})^2  - (q_i^{air})^2}$, $k_p^{air} = n_p^{air} \omega_p / c$,
$k_{i,s}^{air} = n_{i,s}^{air} \omega_{i,s} / c$, $d$ is the distance between the crystals and functions $F_{1}$ and $F_{2}$ are the two-photon amplitudes related to the first and the second crystal, respectively:

\begin{equation} \label{eq:f1_f2}
\begin{aligned}
& F_{1}(\textbf{\textit{k}}_{s}, \textbf{\textit{k}}_{i}) \propto \exp[-\frac{\sigma_{p}^2}{2}(\Delta k_{\parallel}\sin{\rho} + \Delta k_{\perp}\cos{\rho})^2] \textrm{sinc}(\frac{L}{2}\mu) \exp[-i\frac{L}{2}\mu], \\
& F_{2}(\textbf{\textit{k}}_{s}, \textbf{\textit{k}}_{i}) \propto \exp[-\frac{\sigma_{p}^2}{2}(\Delta k_{\parallel}\sin{\rho} + \Delta k_{\perp}\cos{\rho})^2] \textrm{sinc}(\frac{L}{2}\mu)  \exp[i\frac{L}{2}\mu], \\
\end{aligned}
\end{equation}
where $\mu = \Delta k_{\parallel} - \Delta k_{\perp}\tan{\rho}$. Together the whole amplitude can be written as:

\begin{equation} \label{eq:ampl_together}
\begin{aligned}
& F(\textbf{\textit{k}}_{s}, \textbf{\textit{k}}_{i}) \propto \exp[-\frac{\sigma_{p}^2}{2}(\Delta k_{\parallel}\sin{\rho} + \Delta k_{\perp}\cos{\rho})^2] \textrm{sinc}(\frac{L}{2}\mu) \times \\
& \times [\exp(-i\frac{L}{2}\mu) + \exp(i\frac{L}{2}\mu + i\Delta k^{air} d)].\\
\end{aligned}
\end{equation}

For the compensated system shown in Fig.~\ref{fig:2cr_scheme}b, the biphoton amplitude for the second crystal $F_{2}$ can be obtained by changing the anisotropy angle as follows $\rho \rightarrow -\rho$, and is given by:
\begin{equation} \label{eq:f2_compens}
\begin{aligned}
& F_{2}(\textbf{\textit{k}}_{s}, \textbf{\textit{k}}_{i}) \propto \exp[-\frac{\sigma_{p}^2}{2}(-\Delta k_{\parallel}\sin{\rho} - \Delta k_{\perp}\cos{\rho})^2] \textrm{sinc}(\frac{L}{2}\xi)  \exp[i\frac{L}{2}\xi], \\
\end{aligned}
\end{equation}
where $\xi = \Delta k_{\parallel} + \Delta k_{\perp}\tan{\rho}$.

\subsection*{A single crystal at high gain without anisotropy}

For high parametric gain, the properties of light generated in a single crystal without anisotropy can be described through a solution of the following integro-differential equations~\cite{PhysRevResearch.2.013371}:
\begin{eqnarray}
\frac{\mathrm{d} a_s (q_s, z,\omega_s)}{\mathrm{d}z}=
\Gamma \int \mathrm{d}q_i e^{-\frac{(q_s+q_i)^2 \sigma_{p}^2}{2}}  e^{i \Delta k_{\parallel} z} a^{\dagger}_i(q_i, z,\omega_i), \ \ \ \ \
\label{eq:eq1}
\end{eqnarray}
\begin{eqnarray}
\frac{\mathrm{d} a^{\dagger}_i(q_i, z,\omega_i)}{\mathrm{d}z}=
\Gamma \int \mathrm{d}q_s e^{-\frac{(q_s+q_i)^2 \sigma_{p}^2}{2}}  e^{-i \Delta k_{\parallel} z} a_s (q_s, z,\omega_s), \ \ \ \ \
\label{eq:eq2}
\end{eqnarray}
where $\omega_i = \omega_p-\omega_s$, $z \in [0, L]$, the form of the solution is defined in the main text.

\subsection*{A two-crystal system with an air gap and without anisotropy at high gain}

This system of integro-differential equations, describing two crystals without anisotropy with an air gap in-between, has been presented in~\cite{PhysRevResearch.2.013371}. 
Another representation of such a system includes the following expressions for the integro-differential equations in the second crystal:
\begin{eqnarray}
\frac{\mathrm{d} a_s (q_s, z,\omega_s)}{\mathrm{d}z}=
\Gamma \int \mathrm{d}q_i e^{-\frac{(q_s+q_i)^2 \sigma^2}{2}}  e^{i \Delta k_{\parallel} (z + L_{1})} e^{i\Delta k^{air} d} a^{\dagger}_i(q_i, z,\omega_i), \ \ \ \ \
\label{eq1_no_anis_2nd_cr}
\end{eqnarray}
\begin{eqnarray}
\frac{\mathrm{d} a^{\dagger}_i(q_i, z,\omega_i)}{\mathrm{d}z}=
\Gamma \int \mathrm{d}q_s e^{-\frac{(q_s+q_i)^2 \sigma^2}{2}}  e^{-i \Delta k_{\parallel} (z + L_{1})} e^{-i \Delta k^{air} d} a_s (q_s, z,\omega_s), \ \ \ \ \
\label{eq2_no_anis_2nd_cr}
\end{eqnarray}
while for the first crystal Eqs.~(\ref{eq:eq1}) and (\ref{eq:eq2}) are used.

\subsection*{A two-crystal system with an air gap and compensated anisotropy at high gain}

For a system of two crystals in the compensated-anisotropy configuration, the integro-differential equations take the form:
\begin{eqnarray}
\frac{\mathrm{d} a_s (q_s, z,\omega_s)}{\mathrm{d}z}=
\Gamma \int \mathrm{d}q_i \exp{-\frac{\sigma^2}{2}(-\Delta k_{\parallel}\sin{\rho} + \Delta k_{\perp}\cos{\rho})^2 } \times \\ \exp{i \mu L_{1}    + i\Delta k^{air} d + i \xi z} a^{\dagger}_i(q_i, z,\omega_i), \nonumber \ \ \ \ \
\label{eq1_anis_comp_2nd_cr}
\end{eqnarray}
\begin{eqnarray}
\frac{\mathrm{d} a^{\dagger}_i(q_i, z,\omega_i)}{\mathrm{d}z}=
\Gamma \int \mathrm{d}q_s \exp{-\frac{\sigma^2}{2}( -\Delta k_{\parallel}\sin{\rho} + \Delta k_{\perp}\cos{\rho})^2 } \times \\ \exp{-i \mu L_{1}  -i\Delta k^{air} d - i \xi z  } a_s (q_s, z,\omega_s), \nonumber \ \ \ \ \
\label{eq2_anis_comp_2nd_cr}
\end{eqnarray}
where $\xi = \Delta k_{\parallel} + \Delta k_{\perp}\tan{\rho}$.

\end{document}